\documentclass[10pt,letterpaper]{article}

\usepackage{cogsci}

\cogscifinalcopy 

\usepackage{pslatex}
\usepackage{apacite}
\usepackage{float} 

\usepackage{graphicx}
\usepackage{textcomp}
\usepackage{xcolor}
\usepackage{algorithmic}
\usepackage{amsfonts,amssymb}
\usepackage{algorithm}
\usepackage{booktabs}
\usepackage{xspace}
\usepackage{amsmath}
\usepackage{mathabx} 
\usepackage{multirow}
\usepackage{multicol}

\newcommand{\method}{\textit{mix}EEG\xspace}

\title{\method: Enhancing EEG Federated Learning for Cross-subject \\ EEG Classification with Tailored \textit{mixup}}

\author{{\large \bf Xuan-Hao Liu (haogram\_sjtu@sjtu.edu.cn)} \\
  Shanghai Jiao Tong University, Shanghai, China \\
  \AND {\large \bf Bao-Liang Lu (bllu@sjtu.edu.cn)} \\
  Shanghai Jiao Tong University, Shanghai, China \\
  \AND {\large \bf Wei-Long Zheng\thanks{Corresponding author.} (weilong@sjtu.edu.cn)} \\
  Shanghai Jiao Tong University, Shanghai, China
}

\begin{document}

\maketitle



\begin{abstract}
The cross-subject electroencephalography (EEG) classification exhibits great challenges due to the diversity of cognitive processes and physiological structures between different subjects.
Modern EEG models are based on neural networks, demanding a large amount of data to achieve high performance and generalizability.
However, privacy concerns associated with EEG pose significant limitations to data sharing between different hospitals and institutions, resulting in the lack of large dataset for most EEG tasks.
Federated learning (FL) enables multiple decentralized clients to collaboratively train a global model without direct communication of raw data, thus preserving privacy. 
For the first time, we investigate the cross-subject EEG classification in the FL setting.
In this paper, we propose a simple yet effective framework termed \textbf{\method}.
Specifically, we tailor the vanilla \textit{mixup} considering the unique properties of the EEG modality.
\method shares the unlabeled averaged data of the unseen subject rather than simply sharing raw data under the domain adaptation setting, thus better preserving privacy and offering an averaged label as pseudo-label.
Extensive experiments are conducted on an epilepsy detection and an emotion recognition dataset.
The experimental result demonstrates that our \method enhances the transferability of global model for cross-subject EEG classification consistently across different datasets and model architectures. Code is published at: \texttt{\textcolor{blue}{https://github.com/XuanhaoLiu/mixEEG}}.

\end{abstract}
\section{Introduction}
Brain-Computer Interfaces (BCIs) based on electroencephalography (EEG) signals have been significantly facilitating the health care of people in a wide range of areas, encompassing cognitive research \cite{tan2024theta, finley2024resting, liu2024eegvideo}, epilepsy detection \cite{shoeb2010application, tasci2023epilepsy}, and emotion recognition \cite{aggarwal2022review, jiang2023multimodal, jiang2024seed}.
A common but challenging practical requirement of EEG BCI is training an EEG BCI that achieves good performance on unseen subjects, which is also known as cross-subject EEG classification.
Due to the diversity of cognitive processes and physiological structures, electroencephalogram (EEG) data present significant individual differences.
Modern EEG BCIs are largely based on neural networks (NN) and it is widely acknowledged that NN follow the \textit{Scaling Law} \cite{aghajanyan2023scaling, cherti2023reproducible}, i.e., the performance of NN improves when scaling up the amount and quality of training data.
\begin{figure}[htbp]
    \centering
    \includegraphics[width=0.8\linewidth,scale=0.80]{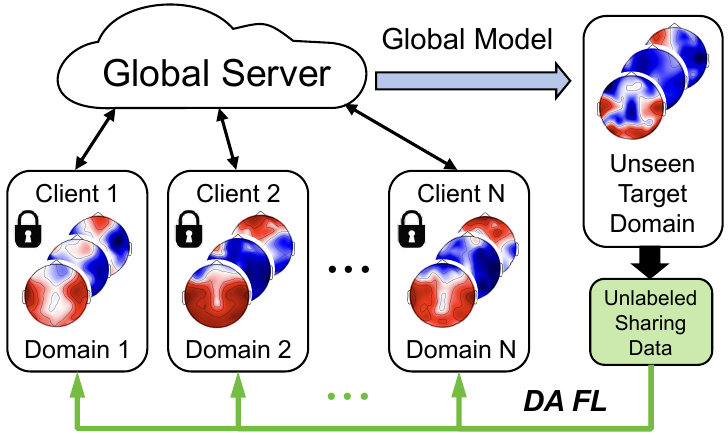}
    \caption{The problem setting of domain generalization federated learning (DG FL) and domain adaptation federated learning (DA FL), which aims to learn a global model from multiple decentralized source domains. The unlabeled sharing data is sent to client only under the DA FL setting.}
    \label{FedLearn}
    \vspace{-0.3cm}
\end{figure}
However, due to the expensive collection costs and privacy concerns \cite{liu2024professor}, there is no public EEG dataset that is large enough to train well generalizable EEG BCIs.
The majority of EEG data around the world is stored in various medical hospitals or research institutes in the format of small datasets.
How can we leverage these dispersed data resources to train more generalizable EEG BCIs?

Federated learning (FL) allows multiple decentralized clients to collaboratively train a global model without direct communication of raw data \cite{liu2024aggregating}.
For instance, during each communication round of the FedAvg \cite{mcmahan2017communication}, the clients update its local model exclusively with their own data, and upload the parameters of their local model to the global server, where the global model is then updated through the aggregation of parameters from these local models.
Although there are many studies adopted FL in EEG-based BCI \cite{ju2020federated, baghersalimi2021personalized, chen2024efficient, chan2024adaptive},
these studies predominantly focused on improving the performance of the global model on existing subjects of each client, overlooking the practical requirement of cross-subject EEG classification on unseen subjects.

Basically, there are two cross-subject settings: domain generalization (DG) and domain adaptation (DA).
While DG requires no data from unseen subjects, DA allows the trainer to access the unlabeled data from unseen subjects.
For the first time, we investigate these two settings in FL, i.e., the DG FL setting and DA FL setting.
Illustrated in Figure~\ref{FedLearn}, these two settings require training a global model on multi-source domains while preserving the data privacy of each source domain.
With the strict prohibition of data communication among source domains, the DG FL and DA FL problems present heightened challenges compared to traditional DG and DA settings, in which the global server can directly access to all source domains data.

Inspired by \textit{mixup} \cite{zhang2018mixup}, a simple yet surprisingly effective data augmentation technique, we propose a novel FL framework called \textbf{\method} to tackle the two proposed problem settings.
Specifically, we tailor the vanilla \textit{mixup} for EEG modality to unleash the potential of \textit{mixup} for developing better cross-subject EEG BCI.
Solely through mixing up the local EEG data of each client, our \method already enhances the generalizability of global models under the DG FL setting.
By adopting the unlabeled data from the target domain, our \method achieves even better performance under the DA FL setting.

In summary, the contributions of this paper are as follows:
\begin{itemize}
    \item \textbf{New problem:} We for the first time investigate two new problem settings for cross-subject EEG FL: the DG FL setting and the DA FL setting. 
    \item \textbf{New framework:} We introduce a new FL framework called \method to improve the transferability of global models trained by FL methods.
    \item \textbf{New mixup:} Beyond the vanilla \textit{mixup}, which linearly interpolates raw data, we specially design and investigate two new \textit{mixup} approaches for EEG data: the \textit{Channel Mixup} and the \textit{Frequency Mixup}.
    \item \textbf{Extensive Experiments:} We conduct extensive experiments on two public datasets and two common network architectures to comprehensively investigate different \method with diverse tailored \textit{mixup}. 
\end{itemize}
\section{Related Work}
\subsection{Cross-subject EEG Classification}
The inter-subject variability of EEG signals has hindered the promotion of EEG BCIs for a long period of time.
Previous DG methods predominantly concentrated on traditional supervised learning or unsupervised learning.
The mixture-of-experts are adopted to learn different brain regions representations \cite{liu2024moge}.
Contrastive learning enables networks to learn the difference among features, which is applied in both a graph-based multi-task self-supervised learning (GMSS) \cite{li2022gmss} and a prototype contrastive domain generalization (PCDG) \cite{cai2023two}.
However, these methods cannot be directly adopted for training a global model with multiple decentralized clients, limiting its access to large amount of data.
Our \method framework enables BCI developer to train a global model with the collaboration of multiple decentralized clients while preserving privacy.

\subsection{EEG Federated Learning}
Considering the privacy preservation of EEG, many previous works introduced FL in EEG classification tasks. 
\citeauthor{liu2024aggregating} proposed FLEEG to to surmount the challengings of the heterogeneity of EEG devices among different clients (hospitals and institutions) \cite{liu2024aggregating}.
FedEEG \cite{hang2023fedEEG} used an inter-subject structure matching-based FL framework to extract the discriminative features for improving the performance of each clients.
\citeauthor{mongardi2024exploring} explored the potential of federated learning for the task of emotion recognition from BCI data, focusing on performance with respect to centralized approaches \cite{mongardi2024exploring}.
Despite there are lots of research on the EEG FL, the cross-subject issues are largely unexplored. Hence, we propose \method to enhance the EEG FL for cross-subject EEG classification.

\subsection{\textit{Mixup}}

\textit{Mixup} \cite{zhang2018mixup} is a simple yet surprisingly effective data augmentation technique, which linearly interpolates two data's inputs and labels to generate a \textit{mixup} data.
By adopting \textit{mixup} strategy, centralized learning is able to train a more robust model with the same training dataset.
Recently, lots of paper utilized \textit{mixup} for fedreated learning in many research fields, like medical image segmentation \cite{wicaksana2022fedmix}, vertical FL \cite{cheng2024fedmix}, and image classification \cite{yoon2021fedmix}.

Previous EEG classification work simply employed the vanilla mixup \cite{zhou2024eegmatch, yao2024advancing}.
To tailor the vanilla \textit{mixup} for diverse data modality, many works like mixing up data in the latent space \cite{verma2019manifold}, detecting the saliency feature for images \cite{uddinsaliencymix}, and \textit{mixup} for natural language \cite{chen2022doublemix} were proposed.
Taking the properties of EEG modality with multiple channels and frequency bands into consideration, we naturally propose \textit{Channel Mixup} and \textit{Frequency Mixup} for EEG modality.
\section{Method}

\subsection{Problem Statement}
Let $\mathcal{D}_s = \{X_s, Y_s\}$ and $\mathcal{D}_t = \{X_t, Y_t\}$ denote the source domain dataset and the target domain dataset. There are two normal cross-subject settings, domain generalization (DG) and domain adaptation (DA). Different from DG setting which requires training DL models only using the source dataset $\mathcal{D}_s$, DA setting allows trainers to have additional access to the unlabeled target data $X_t$.

Under the federated learning setting, there are K clients whose datasets multiple sub-dataset $\mathcal{D}_s^k$, where $k \in 1, ..., K$. For the purpose of privacy preservation, the direct data communication between clients is strictly forbidden, \textit{i.e.}, each client train their local model only using their own sub-dataset $\mathcal{D}_s^k$.
The goal of DG FL is to train a global model that can generalize to $\mathcal{D}_t$.

In contrast to DG FL, DA FL setting allows clients to access the unlabeled data $X_t$ from target domain, making the cross-subject isssue easier but damaging the privacy of target domain $\mathcal{D}_t$.
Hence, it is necessary to consider the trade-off of the privacy preservation and the data sharing.
We define a sharing ratio of $r \in [0, 1]$, denoting the ratio of the number of the shared data to the target domain data.
When $r = 0$, the DA FL setting is equal to the DG FL setting. Larger $r$ represents more data communication but less privacy preservation.

\subsection{Baseline: FedAvg}
Federated Averaging (FedAvg) is the most commonly used algorithm framework in FL \cite{mcmahan2017communication}. For every communication round $t = 0, ..., T-1$, a part of clients train their own local models with their own sub-dataset $\mathcal{D}_s^k$.
We denote the fraction of the number of clients participating in the update to the total number of clients as $\phi$.
Under the DA FL setting, each client also has the access to the unlabeled shared data from target domain dataset $X_t$.
Afterwards, client $k$ sends the model weight $\textbf{\textit{w}}_t^k$ to the global server, which update the parameters of global model by simply averaging the client parameters received by: 
\begin{align}
    \textbf{\textit{w}}_t = \frac{1}{\phi K}\sum_{k=1}^{\phi K} \textbf{\textit{w}}_t^k.
\end{align}
The updated global model $\textbf{\textit{w}}_t$ is sent back to clients for the next round local training. The process will undergoes $T$ times or until the global model converges.

\subsection{Tailored \textit{mixup}}
\begin{figure}[t]
    \centering
    \includegraphics[width=1.0\linewidth,scale=1.0]{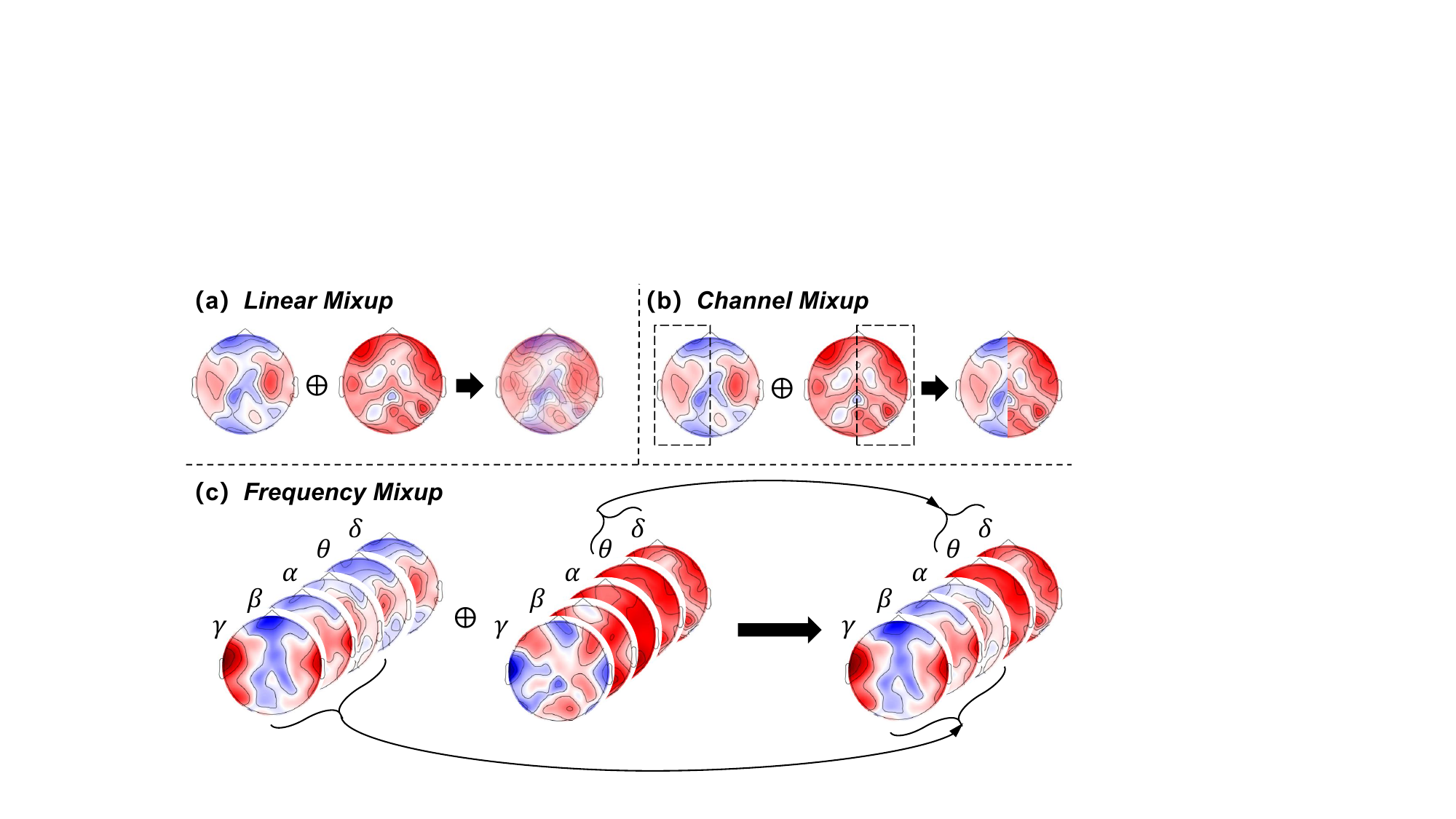}
    \caption{Three EEG mixup methods: (a) Linear Mixup, (b) Channel Mixup: an example that combines the left side and the right side of two EEG signals, (c) Frequency Mixup: an example that that combines the high frequency bands and the low frequency bands of two EEG signals.}
    \label{mixupway}
\end{figure}

\textit{Mixup} \cite{zhang2018mixup} is a simple data augmentation technique using a linear interpolation between two data-label pairs $(\textbf{\textit{x}}_i, y_i)$ and $(\textbf{\textit{x}}_j, y_j)$ to produce a mixup data-label pair
\begin{align}
\widetilde{\textbf{\textit{x}}} = \lambda\textbf{\textit{x}}_i + (1-\lambda)\textbf{\textit{x}}_j,\ \ \
\widetilde{y} = \lambda y_i + (1-\lambda)y_j, 
\end{align}
where $\lambda \in [0,1]$ is a hyperparameter. Through this elegant augmentation, model gains better robustness and generalization \cite{zhang2021does, carratino2022mixup}.
In this paper, we explore and validate three \textit{mixup} methods on DE features as shown in Figure~\ref{mixupway}. It is worth noting that these tailored \textit{mixup} might not always useful as we are the first to explore them.
\begin{itemize}
    \item[1)] \textit{Linear Mixup}: The same as the original mixup.
    \item[2)] \textit{Channel Mixup}: Channel mixup could help the model learn localized spatial representations robust to inter-subject variability in channel activations. There are $C$ channels for EEG data, we separate the whole $C$ channels set into two non-overlapping subsets $C_1$ and $C_2$.
    The mixup data is:
    \begin{align}
        \widetilde{\textbf{\textit{x}}} \in \mathbb{R}^{C\times F} = \textbf{\textit{x}}_i \odot M_r(C_1) + \textbf{\textit{x}}_j \odot M_r(C_2),
    \end{align}
    where $M_r() \in \{0,1\}^{C\times F}$ denotes a binary mask. 
    The rows of $M_r(C_1)$ corresponding to the channels in $C_1$ are 1, and other rows are 0.
    \item[3)] \textit{Frequency Mixup}: Frequency mixup may enable capturing discriminative features across multiple spectral bands, which is known to be important for EEG analysis tasks. There are $F$ frequency bands for EEG data, we separate the whole $F$ frequency bands set into two non-overlapping subsets $F_1$ and $F_2$.
    The mixup data is:
    \begin{align}
        \widetilde{\textbf{\textit{x}}} \in \mathbb{R}^{C\times F} = \textbf{\textit{x}}_i \odot M_c(F_1) + \textbf{\textit{x}}_j \odot M_c(F_2), 
    \end{align}
    where $M_c() \in \{0,1\}^{C\times F}$ denotes a binary mask.
    The columns of $M_c(F_1)$ corresponding to the frequency bands in $F_1$ are 1, and other columns are 0.
\end{itemize}

\subsection{\method}
For FedAvg algorithm framework, the local clients updates the client model via stochastic gradient descent (SGD) for $E$ epochs:
\begin{align}
    \textbf{\textit{w}}_{t+1, i+1}^k \leftarrow \textbf{\textit{w}}_{t+1, i}^k - \eta \nabla \mathcal{L}(f(\textbf{\textit{x}}_i^k;\textbf{\textit{\textit{w}}}_{t+1, i}), y_i^k), i = 0, ..., E-1.
\end{align}
Here, $\eta$ is learning rate, $\mathcal{L}$ is loss function, and $f(\textbf{\textit{x}}, \textbf{\textit{w}})$ is the model output of the input
$\textbf{\textit{x}}$ given model weight $\textbf{\textit{w}}$.

\begin{algorithm}[h]
	\caption{\method under the DA FL setting}
	\label{alg:FAAM}
	\begin{algorithmic}[1]
            \STATE{\textbf{Input:} $\mathcal{D}_s^k = {\textbf{\textit{X}}_s^k, \textbf{\textit{Y}}_s^k}$ for $k = 1, ..., N$, $\mathcal{D}_t$, sharing ratio $r \in [0, 1]$, when $r=0$, this algorithm is FedAvg.}
		\STATE {Initialize $\textbf{\textit{w}}_0$ for global model.}
		\STATE {$\textbf{\textit{X}}_g \leftarrow GetSharingData(\mathcal{D}_t, r)$} 
            \FOR{$t = 0, ..., T-1$}
                \FOR{$k = 1, ..., N$}
                    \STATE {$\textbf{\textit{w}}_{t+1}^k \leftarrow LocalUpdate(k, \textbf{\textit{w}}_t; \textbf{\textit{X}}_g)$}
                \ENDFOR
                \STATE {$\textbf{\textit{w}}_t \leftarrow \frac{1}{N}\sum_{k=1}^N \textbf{\textit{w}}_t^k$}
            \ENDFOR
	\end{algorithmic}
\end{algorithm}

\begin{algorithm}[h]
	\caption{Local Update for \method}
	\label{alg:LocalUpdate}
	\begin{algorithmic}[1]
            \STATE{\textbf{Input:} $k, \textbf{\textit{w}}_t; \textbf{\textit{X}}_g$, when $\textbf{\textit{X}}_g = \varnothing$, $\mathcal{L}_2 = 0$}
		\STATE {$\textbf{\textit{w}} \leftarrow \textbf{\textit{w}}_t$} 
            \FOR{$e = 0, ..., E-1$}
                \FOR{$batch(\textbf{\textit{X}}, \textbf{\textit{Y}})$}
                    \STATE {Select a sharing data $\textbf{\textit{x}}_g$ from $\textbf{\textit{X}}_g$}
                    \STATE {$\textbf{\textit{X}}_m \leftarrow$ mixup \textbf{\textit{X}} with $\textbf{\textit{x}}_g$}
                    \STATE {$\mathcal{L}_1 \leftarrow \mathcal{L}(f(\textbf{\textit{X}}_m; \textbf{\textit{w}}), \textbf{\textit{Y}})$}
                    \STATE {$\mathcal{L}_2 \leftarrow \sum_{c=0}^{n\_class} \mathcal{L}(f(\textbf{\textit{X}}_m; \textbf{\textit{w}}), \mathsf{One}(c))$}
                    \STATE {$\textbf{\textit{w}} \leftarrow \textbf{\textit{w}} - \eta_{t+1} \nabla (\lambda\mathcal{L}_1 + (1-\lambda)\mathcal{L}_2)$}
                \ENDFOR
            \ENDFOR
            \RETURN{\textbf{\textit{w}}}
	\end{algorithmic}
\end{algorithm}

The SGD algorithm for updating local models usually leads these models to overfit their own subset, damaging the global model's generalizability and making the global model hard to do cross-subject EEG classification.
To this end, we propose \method, a simple but effective framework for enhancing the generalizability of the global EEG model learned under the FL setting.

For the DG FL setting, \method simply replacing the local updating strategy of local clients from SGD with \textit{mixup}, i.e. using \textit{mixup} technique to update the local model $\textbf{\textit{w}}_t^k$. We validate several different \textit{mixup} strategies to fully investigate which type of \textit{mixup} is suitable for EEG modality.

For the DA FL setting, local clients are able to see the $r\cdot |\mathcal{D}_t|$ unlabeled data.
Here are two questions: \textbf{Q1}: what is the shared data? \textbf{Q2}: how do local clients leverage these unlabeled data?

For question \textbf{Q1}, we argue that directly sharing the subset of $X_t$ definitely harm the privacy security of the target domain.
So instead, we propose to generate shared data by averaging the target domain data $X_t$.
Specifcally, we define a hyperparameter $s \in \mathbb{Z}^+$, indicating the number of data used for aggregating a single averaged data.
We randomly shuffle the $X_t$, and linearly interpolating a number of $s$ data from $X_t$ to generate a single shared data by Eq \ref{eq:gen}. This process is repeated $r\cdot |\mathcal{D}_t|$ times to get enough shared data.
\begin{align}
    \widetilde{\textbf{\textit{x}}_t} = \frac{1}{s}\sum_{i=1}^{s}{\textbf{\textit{x}}_t^i}.
    \label{eq:gen}
\end{align}

For question \textbf{Q2}, by assuming that the appearing probability of each emotion in the target dataset is equal, we can generate a pseudo label for each shared data: $\tilde{y}_t = \frac{1}{c}[1,1,...,1] \in \mathbb{R}^{c}$,
where $c$ is the number of emotion classes.
After generating the shared data and correpsonding labels, \method can leverage these data from target dataset to improve the transfer ability.
Unlike the DG FL issue, where we mixup data within a single batch, we mixup each data-label pairs $(\textbf{\textit{x}}_s^i, y_s^i)$ from $\mathcal{D}_s^k$ with the shared data. Decoding $\mathsf{One}()$ as the onehot encoding, the corresponding mixup label is:
\begin{align}
    \widetilde{y} = \lambda \mathsf{One}(y_s^i) + (1-\lambda)\frac{1}{c}[1,1,...,1].
\end{align}
\section{Experiments}
\subsection{EEG Datasets}
We exploit two datasets: an epilepsy detection dataset \cite{shoeb2010application} and an emotion recognition dataset \cite{zheng2015investigating}. Differential Entropy (DE) features \cite{duan2013differential} are extracted for classification.

\textbf{Epilepsy Detection (ED)}: CHB-MIT dataset is an epilepsy dataset required from 23 patients. We cropped and resampled the CHB-MIT dataset to build an ED dataset with four types of EEG: ictal, preictal, postictal, and interictal phase EEG.

\textbf{Emotion Recognition (ER)}: SEED dataset records the 62-channels EEG data corresponding to 3 types of emotions from 15 subjects: positive, neutral, and negative.
\subsection{Experimental Setup}
Each subject is regarded as a decentralized source domain, and the results under leave-one-subject-out (LOSO) settings are reported.
Specifically, we select $n-1$ subjects as source domains, train a global model using FL methods on the source domains (the direct data communication among source domains) and test the performance on the last remaining subject, which is regarded as target domain.

The number of update round of the global server $T = 50$, for each client, the number of local update round $E = 5$. We use SGD as the optimizer, with its learning rate $\eta = 0.01$. The fraction $\phi$ is set to 0.2 for ER dataset, and 0.3 for ED dataset. The sharing ratio $r$ is 0.1 and aggregating hyperparameter $s=10$ for DA FL setting.

\subsection{EEG Models}
There are many methods for the classification of EEG data. We choose the following two type of models: \textbf{1) MLP:} Multi-Layer Perception, here we use an MLP with 2 hidden layers of 128 and 64 neurons. \textbf{2) CNN:} Convolutional Neural Network, here we use a 3 layer CNN with kernal sizes of \{(3,3), (3,3), and (7,1)\}. The activate function is the GELU function and the dropout probability is 0.5 for all models.

\begin{table*}[htbp]
	\centering
	\caption{\textbf{The DG FL setting}: The accuracy, F1 score and Cohen's Kappa score (\%) under the DG FL setting where we set $r = 0.1$ in our experiment. L, C, and F indicate Linear, Channel, and Frequency mixup strategies, respectively.}
        {\small
	\begin{tabular}{llcccccc} 
		\toprule
		~ & ~ &\multicolumn{3}{c}{\textbf{Epilepsy Detection: CHB-MIT, 4-class}}&\multicolumn{3}{c}{\textbf{Emotion Recognition: SEED, 3-class}}\\
		\cmidrule(lr){3-5}\cmidrule(lr){6-8}
		~ & \textbf{Methods}&\textbf{Accuracy}&\textbf{F1 Score}&\textbf{Cohen's Kappa }&\textbf{Accuracy}&\textbf{F1 Score}&\textbf{Cohen's Kappa}\\
            \midrule
		\multirow{7}*{\rotatebox{90}{MLP}} & FedAvg &43.676 $\pm$ 4.367& 41.763 $\pm$ 4.349&27.376 $\pm$ 5.823 & 77.053 $\pm$7.824 & 76.826 $\pm$ 8.008 & 65.514 $\pm$ 11.768 \\
		~ & L $\alpha$ = 0.2 &47.618 $\pm$ 3.925&\textbf{46.895 $\pm$ 3.532}&30.158 $\pm$ 5.233 & 70.505 $\pm$ 4.987 & 69.946 $\pm$ 5.252 &55.625 $\pm$ 7.551 \\
		~ & L $\alpha$ = 5 &\textbf{47.679 $\pm$ 4.064}&46.103 $\pm$ 3.303&\textbf{30.238 $\pm$ 5.419} & 77.387 $\pm$ 8.216 &76.856 $\pm$ 8.536 & 65.939 $\pm$ 12.409 \\
		~ & C \textit{Binary} &47.742 $\pm$ 3.863&45.766 $\pm$ 3.365&30.322 $\pm$ 5.151 & 86.082 $\pm$ 4.607 & 85.936 $\pm$ 4.753 & 79.062 $\pm$ 6.924 \\
		~ & C \textit{Random} &46.984 $\pm$ 3.907&44.888 $\pm$ 3.139&29.312 $\pm$ 5.209 & \textbf{86.306 $\pm$ 6.158} &\textbf{86.148 $\pm$ 6.246} &\textbf{79.419 $\pm$ 9.232} \\
		~ & F $\{\alpha, \beta, \gamma\}$ &47.359 $\pm$ 3.773&44.941 $\pm$ 3.353&29.812 $\pm$ 5.031 & 83.079 $\pm$ 3.562 & 82.910 $\pm$ 3.594 & 74.524 $\pm$ 5.383 \\
		~ & F $\{\delta, \alpha, \gamma\}$ &45.863 $\pm$ 4.359&44.432 $\pm$ 3.612&27.818 $\pm$ 5.812 & 82.224 $\pm$ 4.200 & 82.037 $\pm$ 4.249 & 73.240 $\pm$ 6.276 \\
        \midrule
		\multirow{7}*{\rotatebox{90}{CNN}} & FedAvg &45.142 $\pm$ 4.420&44.000 $\pm$ 4.732&26.857 $\pm$ 5.894 & 83.943 $\pm$ 7.437 & 83.843 $\pm$ 7.443 & 75.876 $\pm$ 11.143 \\
		~ & L $\alpha$ = 0.2 &\textbf{48.606 $\pm$ 4.266}&\textbf{48.047 $\pm$ 4.017}&\textbf{31.474 $\pm$ 5.688} & \textbf{88.814 $\pm$ 4.928} &\textbf{88.759 $\pm$ 4.953} &\textbf{83.205 $\pm$ 7.396} \\
		~ & L $\alpha$ = 5 &46.698 $\pm$ 3.275&44.967 $\pm$ 2.530&28.931 $\pm$ 4.367 & 87.254 $\pm$ 5.789 & 87.118 $\pm$ 4.706 &80.850 $\pm$ 8.694 \\
		~ & C \textit{Binary} &47.127 $\pm$ 4.147&45.587 $\pm$ 4.104&29.503 $\pm$ 5.530 & 84.486 $\pm$ 6.368 &84.399 $\pm$ 6.384 &76.718 $\pm$ 9.542 \\
		~ & C \textit{Random} &46.577 $\pm$ 3.782&43.526 $\pm$ 3.925&28.771 $\pm$ 5.042 & 85.989 $\pm$ 7.578 &85.916 $\pm$ 7.598 &79.007 $\pm$ 11.301 \\
		~ & F $\{\alpha, \beta, \gamma\}$ &46.246 $\pm$ 3.467&44.000 $\pm$ 3.358&28.328 $\pm$ 4.622 & 86.890 $\pm$ 5.445 & 86.818 $\pm$ 5.549 & 80.304 $\pm$ 8.180 \\
		~ & F $\{\delta, \alpha, \gamma\}$ &45.338 $\pm$ 3.251&43.599 $\pm$ 2.942&27.118 $\pm$ 4.335 & 87.170 $\pm$ 5.045 & 87.075 $\pm$ 5.124 & 80.732 $\pm$ 7.571 \\
		\bottomrule
	\end{tabular}}
	\label{tab:DGFL}
\end{table*}

\subsection{EEG Mixup Appoaches}
Basically, we evaluate three \textit{mixup} ways, but there are more implementation details to evaluate: \textbf{1) \textit{Linear Mixup}}: In the origin paper, $\lambda \sim$ Beta$(\alpha, \alpha), \alpha \in (0, +\infty)$, we test two cases: $\alpha = 0.2$ and $\alpha = 5$. \textbf{2) \textit{Channel Mixup}}: We test two cases: (1) dividing the EEG channels into the left scalp and the right scalp; (2) dividing the EEG channels equally into two sets randomly. We set $\lambda=0.5$ for both cases. \textbf{3) \textit{Frequency Mixup}}: For the DE features with five frequency bands ($\delta$, $\theta$, $\alpha$, $\beta$, and $\gamma$), we test two cases: (1) dividing the EEG frequency bands into the $\{\alpha, \beta, \gamma\}$ and $\{\delta, \theta\}$; (2) dividing the EEG frequency bands into the $\{\delta, \alpha, \gamma\}$ and $\{\theta, \beta\}$. We set $\lambda=0.6$ for both cases.
\section{Results}
\subsection{DG FL Setting}
Under DG FL settings, all clients can only access to their own data. We compare our \method with the baseline FedAvg method. The LOSO results are presented in Table~\ref{tab:DGFL}.
It can be observed that simply updating the client models with local mixup have already enhanced the transferability of global model, as most results of \method are higher than the results of the baseline FedAvg. \method improves the cross-subject EEG classification from 77.0\% to 86.3\% for ER task, and from 43.6\% to 47.6\% for ED task.

It can be seen that for MLP architecture, \textit{Linear} \textit{Mixup} is the best for ED task, while \textit{Channel} \textit{Mixup} with Binary Split is the best for ER task. However, for CNN architecture, \textit{Linear} \textit{Mixup} is always better across two datasets.

\subsection{DA FL Setting}
Under DA FL settings, there are $r \cdot |\mathcal{D}_t|$
unlabeled data from target domain sharing to local clients.
The results are presented in Table \ref{tab:DAFL}, it can been observed that \method enhances the performance from 46\% to 49\% for the ED task and from 90\% to 93\% for the ER task.

In this setting, the \textit{Channel Mixup} method with Binary Split outperforms other \textit{mixup} strategies for MLP architecture across two datasets.
While for CNN architecture, the \textit{Linear Mixup} exhibits the best improvement on these two tasks.
However, we notice that the performance of \textit{Frequency Mixup} doesn't work well under the DA FL setting, except for MLP architecture on the ED dataset, \textit{Frequency Mixup} always harms the generalizabitliy for other cases.
Taking the performance of the DG FL setting into consideration, we can conclude that \textit{Frequency Mixup} might not be suitable for EEG modality.

\begin{table*}[h]
	\centering
	\caption{\textbf{The DA FL setting}: The accuracy, F1 score and Cohen's Kappa score (\%) under the DA FL setting. L, C, and F indicate Linear, Channel, and Frequency \textit{mixup} strategies, respectively.}
        {\small
	\begin{tabular}{llcccccc} 
		\toprule
		~ & ~ &\multicolumn{3}{c}{\textbf{Epilepsy Detection: CHB-MIT, 4-class}}&\multicolumn{3}{c}{\textbf{Emotion Recognition: SEED, 3-class}}\\
		\cmidrule(lr){3-5}\cmidrule(lr){6-8}
		~ & \textbf{Methods}&\textbf{Accuracy}&\textbf{F1 Score}&\textbf{Cohen's Kappa}&\textbf{Accuracy}&\textbf{F1 Score}&\textbf{Cohen's Kappa}\\
            \midrule
		\multirow{7}*{\rotatebox{90}{MLP}} & FedAvg &46.806 $\pm$ 4.730&45.948 $\pm$ 4.989&30.742 $\pm$ 6.307 & 90.832 $\pm$ 6.181 &90.769 $\pm$ 6.266 &86.222 $\pm$ 9.290 \\
		~ & L $\alpha$ = 0.2 &48.572 $\pm$ 4.437&47.065 $\pm$ 4.331&31.429 $\pm$ 5.919 & 91.126 $\pm$ 5.376 &91.007 $\pm$ 5.521 & 86.642 $\pm$ 8.099 \\
		~ & L $\alpha$ = 5.0 &47.543 $\pm$ 3.945&45.040 $\pm$ 4.008&30.058 $\pm$ 5.260 & 90.933 $\pm$ 5.808 &90.906 $\pm$ 5.814 &86.379 $\pm$ 8.714 \\
		~ & C \textit{Binary} &\textbf{49.181 $\pm$ 4.558}&\textbf{47.936 $\pm$ 4.208}& \textbf{32.242 $\pm$ 6.077} & \textbf{93.077 $\pm$ 3.690} &\textbf{93.046 $\pm$ 3.702} &\textbf{89.594 $\pm$ 5.543} \\
		~ & C \textit{Random} &48.625 $\pm$ 4.103&47.649 $\pm$ 3.733&31.500 $\pm$ 5.471 & 89.770 $\pm$ 4.386 &89.616 $\pm$ 4.535 &84.624 $\pm$ 6.568 \\
		~ & F $\{\alpha, \beta, \gamma\}$ &48.305 $\pm$ 4.351&47.237 $\pm$ 4.053& 31.074 $\pm$ 5.802 &88.602 $\pm$ 4.656 &88.466 $\pm$ 4.730 &82.865 $\pm$ 6.996 \\
		~ & F $\{\delta, \alpha, \gamma\}$ &47.533 $\pm$ 3.979&45.944 $\pm$ 3.614&30.045 $\pm$ 5.306 & 89.255 $\pm$ 5.929 &89.174 $\pm$ 5.969 &83.878 $\pm$ 8.868 \\
        \midrule
		\multirow{7}*{\rotatebox{90}{CNN}} & FedAvg &48.431 $\pm$ 3.978&\textbf{47.802 $\pm$ 3.496}&31.242 $\pm$ 5.304 & 90.013 $\pm$ 4.722 &89.997 $\pm$ 4.699 & 84.998 $\pm$ 7.086 \\
		~ & L $\alpha$ = 0.2 &\textbf{48.484 $\pm$ 3.773}&47.661 $\pm$ 3.473&\textbf{31.313 $\pm$ 5.031} & 90.671 $\pm$ 5.113 &90.654 $\pm$ 5.097 &85.966 $\pm$ 7.689 \\
		~ & L $\alpha$ = 5 &47.686 $\pm$ 4.068&45.914 $\pm$ 3.357&30.248 $\pm$ 5.425 & \textbf{92.795 $\pm$ 4.469} &\textbf{92.733 $\pm$ 4.552} & \textbf{89.160 $\pm$ 6.734} \\
		~ & C \textit{Binary} &47.427 $\pm$ 4.277&46.233 $\pm$ 4.143&29.903 $\pm$ 5.702 & 88.784 $\pm$ 5.675 &88.753 $\pm$ 5.701 &83.142 $\pm$ 8.529 \\
		~ & C \textit{Random} &47.654 $\pm$ 3.497&46.588 $\pm$ 3.230&30.206 $\pm$ 4.662 & 89.796 $\pm$ 5.265 &89.705 $\pm$ 5.346 &84.675 $\pm$ 4.706 \\
		~ & F $\{\alpha, \beta, \gamma\}$ &47.529 $\pm$ 3.690& 46.519 $\pm$ 3.145&30.038 $\pm$ 4.920 & 88.391 $\pm$ 6.627 &88.272 $\pm$ 6.779 &82.549 $\pm$ 9.952 \\
		~ & F $\{\delta, \alpha, \gamma\}$ &47.330 $\pm$ 3.907&46.507 $\pm$ 3.654&29.774 $\pm$ 5.209 & 88.366 $\pm$ 8.192 &88.261 $\pm$ 8.253 &82.491 $\pm$ 12.334 \\
		\bottomrule
	\end{tabular}}
	\label{tab:DAFL}
\end{table*}

\subsection{Ablation Study}
\subsubsection{\textit{Linear} \textit{Mixup} Hyperparameter $\alpha$}
\begin{figure}[H]
    \begin{center}
    \includegraphics[width=0.9\linewidth]{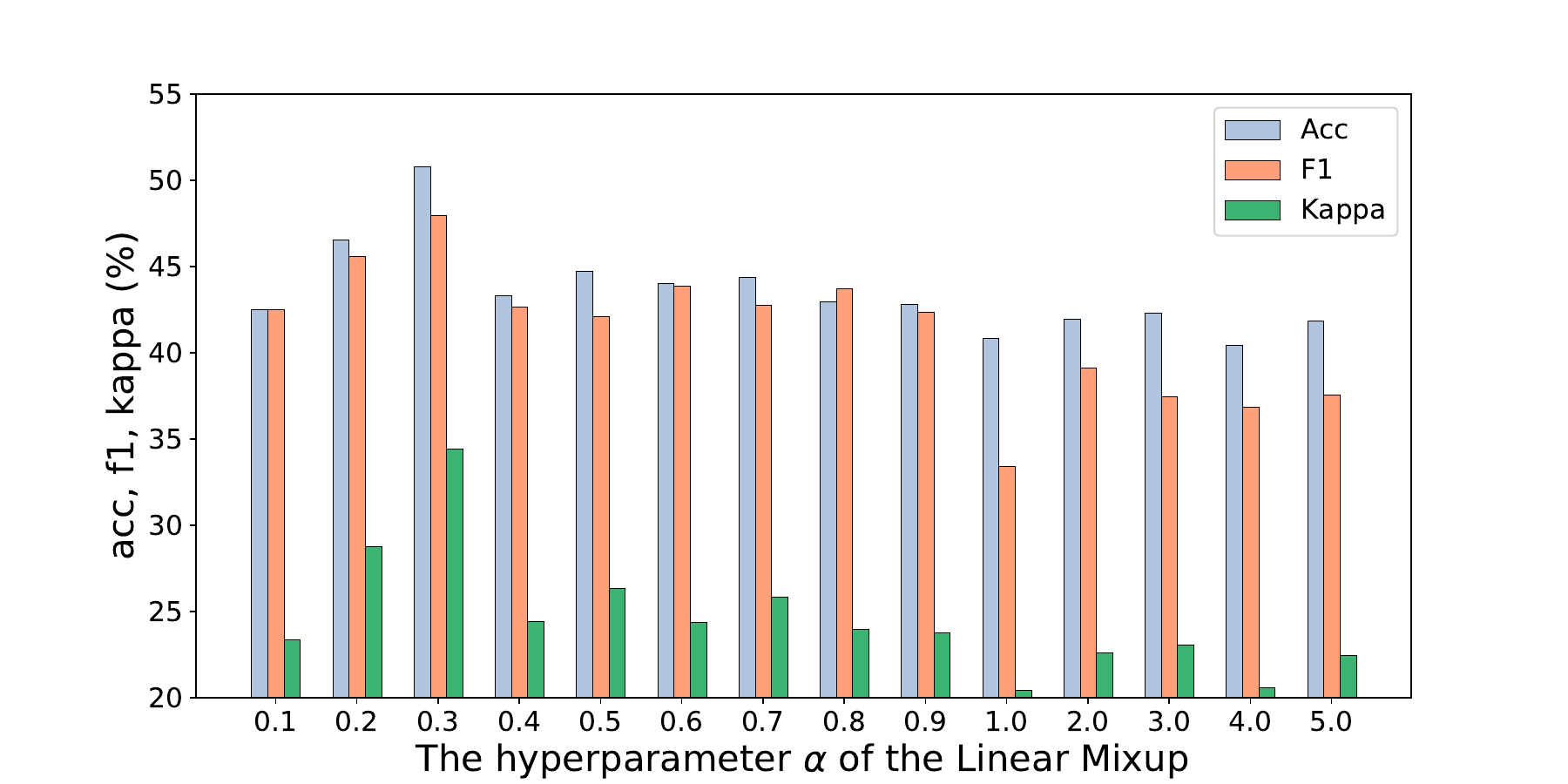}
    \end{center}
    \vspace{-3mm}
    \caption{The accuracy, F1 score and Cohen’s Kappa score of \textit{Linear} \textit{Mixup} with various $\alpha$ on the Epilepsy Detection dataset. The network architecture is MLP.} 
    \label{fig:abl_alpha}
\end{figure}

Through the experiments of the DG/DA FL settings, it can be found that the performance of \textit{Linear Mixup} are influenced by the hyperparameter $\alpha$. Thus, we conduct an extra experiment to evaluate the impact of $\alpha$. We choose 14 values for $\alpha$ and exhibits the performance of CNN model on the ED dataset, as shown in Fig \ref{fig:abl_alpha}. It can be seen that in this case, the best performance is achieved when $\alpha$ is around 0.3. When $\alpha$ is larger than 1.0, the performances are similar.
\vspace{-3mm}
\begin{figure}[H]
    \vspace{-3mm}
    \begin{center}
    \includegraphics[width=0.9\linewidth]{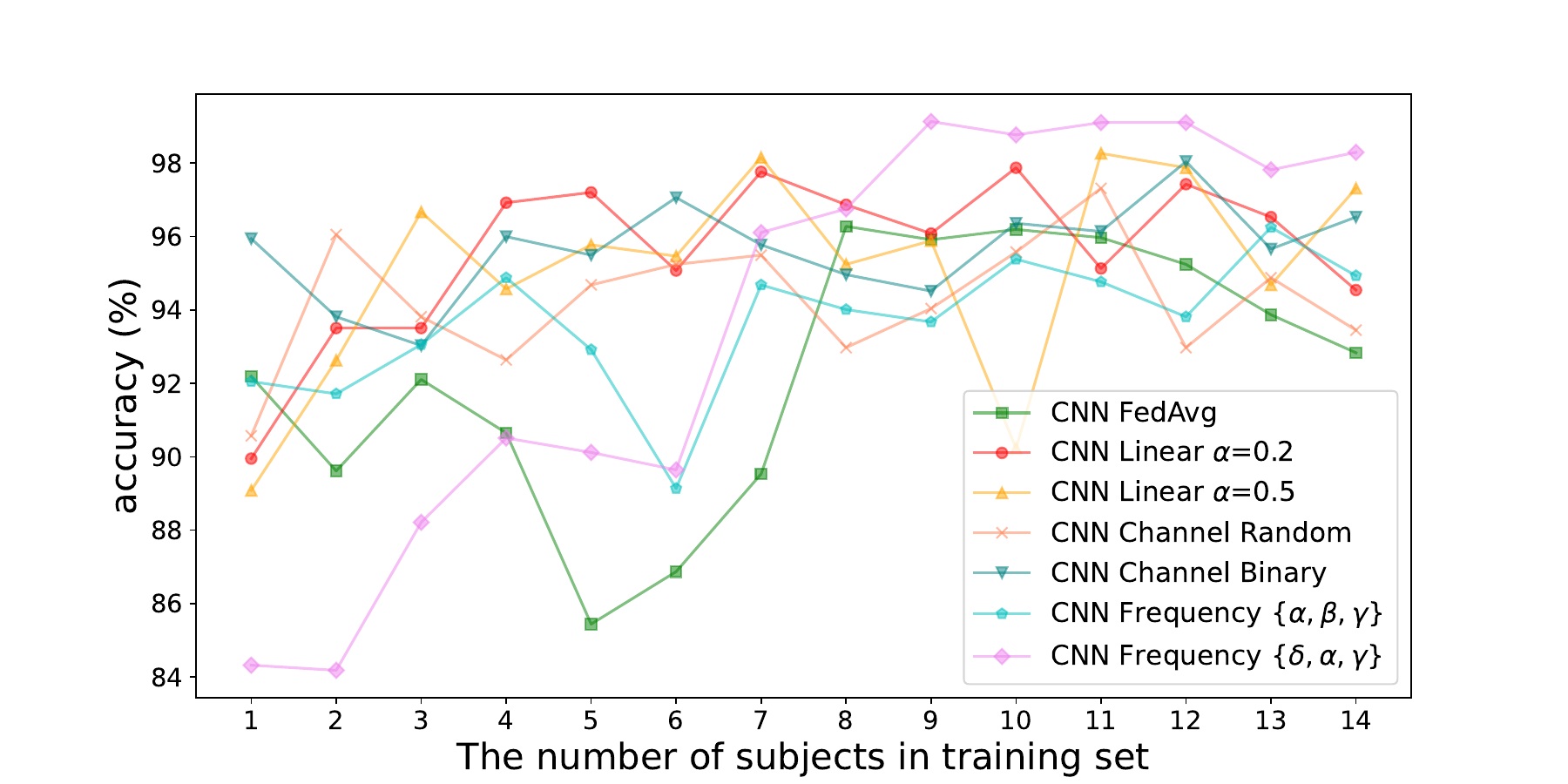}
    \end{center}
    \vspace{-3mm}
    \caption{The accuracy of the global model trained with different training set sizes on the SEED dataset. We choose the No.15 subject as the target domain, and adopt other 14 subjects as the training set. The network architecture is CNN.} 
    \label{fig:abl_size}
\end{figure}
\vspace{-3mm}

\subsubsection{Training Set Size}
To investigate the \textit{Scaling Law} in EEG BCIs, we further train the global model with different numbers of source domain clients.
As depicted in Fig \ref{fig:abl_size}, for all \textit{mixup} strategies, the performance fluctuates since the the EEG data from different subjects vary significantly.
But despite the fluctuation, the performances increase when the training set get larger, demonstrating that larger training data helps enhancing the generalizabitliy.
{\begin{figure}
    \begin{center}
    \includegraphics[width=0.9\linewidth]{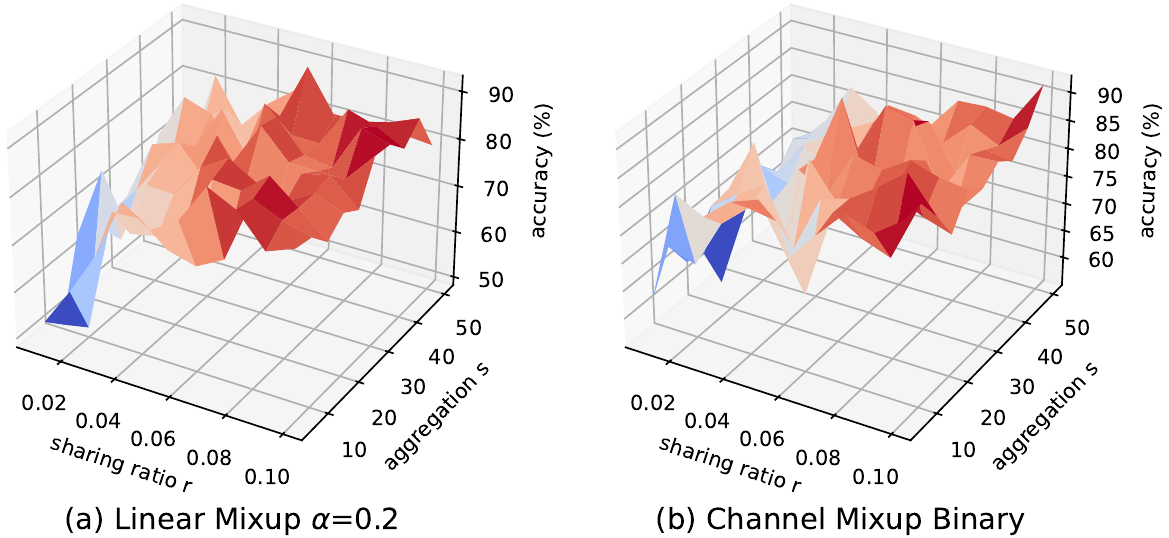}
    \end{center}
    \vspace{-3mm}
    \caption{The 3D heatmap of the LOSO results on the SEED dataset with different sharing ratio $r$ and aggregating number $s$. The network architecture is MLP.}
    \setlength{\belowcaptionskip}{-3cm}
    \label{fig:3d}
    \vspace{-3mm}
\end{figure}

}

\subsubsection{Sharing ratio and Aggregating number}

We conduct additional experiments to validate the effectiveness of the sharing data by adjusting the sharing ratio $r$ and the aggregating hyperparameter $s$. The range of $r$ is from 0.01 to 0.1 and the range of $s$ is from 5 to 50 (step is 5). Since this experiment is quite large (consists of 100 sets of hyperparameters), we only run the global model for 10 epochs, i.e., $T = 10$ in this case.
Displayed in Fig \ref{fig:3d}, it can be observed that the performances increase when sharing ratio $r$ increases, indicating that more sharing data helps cross-subject EEG classification.
We can also see that though the performances fluctuates, larger aggregation $s$ achieves better results.

\subsection{Representation Visualization}
To better present the difference between each \textit{mixup} strategies and explore the reason why some \textit{mixup} strategies are better on learning transferable features, we adopt UMAP technique \cite{mcinnes2018umap} to visualize the outputs of the global model's last layer, \textit{i.e.}, the learned features of the target domain's data.

\begin{figure}
    \begin{center}
    \includegraphics[width=1\linewidth]{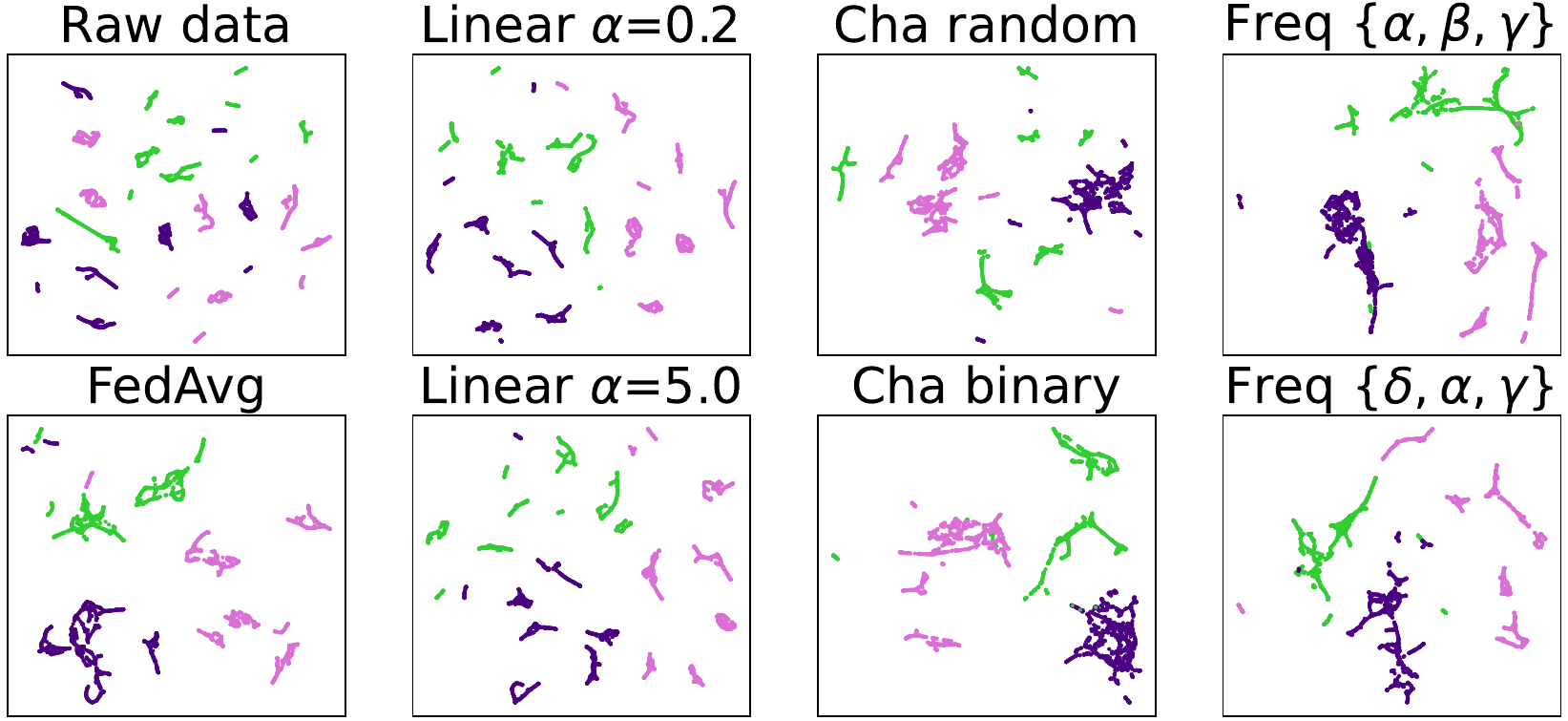}
    \end{center}
    \vspace{-3mm}
    \caption{The UMAP visualization of learned feature on the SEED dataset of each global model using different training strategy under the DA FL setting. Purple, Green, and Pink stands for the negative, neutral, and positive emotion.} 
    \setlength{\belowcaptionskip}{-3cm}
    \label{fig:vis}
\end{figure}

Demonstrated in Fig \ref{fig:vis}, where we plot the learned feature of the MLP architechture on the SEED dataset.
It can be seen that FedAvg's feature has some chaotic distribution on the left up side.
\textit{Linear Mixup} tends to learn more dispersed features,
thus may increase the generalizability in some cases.
In this case (MLP on ER task), the \textit{Channel Mixup} with Binary is the best, it can be seen that the learned feature are more distinguishable since the features of the same class are aggregated more tightly.
\textit{Frequency Mixup} also enables the global model to learn more tightly aggregated features, but the learning on the boundaries between classes is relatively vague, leading to suboptimal generalizabitliy.

\vspace{-0.1cm}
\section{Conclusion}
In this paper, we investigate two new problem settings in EEG-based BCI, the DG FL setting and the DA FL setting. We propose a simple but effective framework \method, and investigate multiple tailored \textit{mixup} strategies on two datasets. We found that by sharing averaged data, \method can leverage these unlabeled data to further improve the generalizabitliy of global model. Extensive experiments validate the efficacy of \method and indicate that \textit{Linear Mixup} (CNN perfers) and \textit{Channel Mixup} (MLP perfers) are useful for enhancing generalizability.


\bibliographystyle{apacite}

\setlength{\bibleftmargin}{.125in}
\setlength{\bibindent}{-\bibleftmargin}

\newpage
\bibliography{ref}

\end{document}